\documentstyle[prb,multicol,aps,epsf]{revtex}

\begin{document}
\draft

\title{Numerical Method for  Zero-Temperature Vortex-Line Phase Diagrams}    

\author{ Welles A. M. Morgado and Gilson Carneiro
\footnote{Corresponding author. Fax: 55 21 290 9020. e-mail: gmc@if.ufrj.br}}
\address{Instituto de F\'{\i}sica\\Universidade Federal do Rio de Janeiro\\  
C.P. 68528\\ 21945-970, Rio de Janeiro-RJ \\ Brazil }
\date{\today}
\maketitle
\begin{abstract}
A numerical method to calculate equilibrium vortex-line configurations in
bulk anisotropic type-II superconductors, at zero temperature, placed in
an external magnetic field is introduced and applied to two physical
problems. The method is designed to search for the minimum of the Gibbs
free-energy in the  London approximation and  assumes only that the vortex 
lines are straight and arranged in a periodic lattice. Based on these
assumptions a simulated annealing algorithm is developed to find the
vortex-line-lattice unit cell shape, and vortex-line  positions within
it. This algorithm is made fast and accurate by the use of a rapidly
converging series to calculate the lattice sums entering the
vortex-vortex interaction energy. The method's accuracy is illustrated
by calculating the magnetic induction versus applied field curve for an
isotropic superconductor. The method  is applied to a superconductor
with a square lattice of columnar defects to study selected regions of the
zero-temperature-phase diagram where  vortex-line-lattices
commensurate with the columnar defect lattice exist.

\end{abstract}

\begin{multicols}{2}
\narrowtext
%\widetext

\section{Introduction} 
\label{sec.int}

A  first step for  understanding the behavior of type-II
superconducting systems is to  determine the 
equilibrium vortex configurations that occur at temperature zero, or
the zero-temperature phase diagram. Although this problem has been
under study for many years \cite{revs}, there are many  systems, such as
artificial structures, constrained geometries and anisotropic
superconductors, for which some questions remain open.

The situation  of greater interest is that of a superconductor
subjected to an external  magnetic field ${\bf H}$. The theoretical
problem of calculating the zero-temperature phase diagram is  then to
find the vortex configuration that minimizes the Gibbs free-energy  for
a given ${\bf H}$. The analytical method to solve this problem
determines, instead,  the  vortex configurations that minimize the
energy for a given  magnetic induction, ${\bf B}$, and then 
calculates ${\bf H}$ from the derivatives of the energy with respect to
${\bf B}$ \cite{revs}. Analytical solutions are difficult to obtain and
are known only for a few  situations. Numerical methods to solve this problem
have been proposed. These  attempt to find the zero-temperature phase diagram 
of a large number of vortices by minimizing the energy, if ${\bf B}$ is
kept constant, or the Gibbs free-energy, if ${\bf H}$ s kept
constant\cite{bro,nor}. These methods are  severely limited by
finite-size effects.

In this paper we introduce a numerical method  to calculate 
configurations of straight vortex lines that minimize the Gibbs
free-energy of bulk superconductors in
the London approximation for a given ${\bf H}$. The
essential ingredients of our methods are two. First, we assume that the
vortex lines are straight and restrict the 
search for the minimum free-energy configurations  to periodic
vortex-line-lattices  (VLL) with a given number of vortex lines per unit cell.
This restriction greatly reduces the number of parameters that we have
to vary to find the minimum. Second, we use a rapidly converging
series  to calculate the lattice sums entering the vortex-vortex 
interaction energy in the London approximation \cite{mmd}. These
ingredients allow us to create a simulated annealing  algorithm, using
Monte Carlo techniques, that
is fast and, as a consequence, highly accurate. A closely related 
method has been used in Ref.\cite{gmc}  to calculate the
zero-temperature phase diagram for films under parallel fields.

To illustrate the accuracy of the method we apply it to the well known
problem of a bulk superconductor under a field $H$ to obtain the
$B\times H$ curve and to compare it with known results at high and low fields
\cite{parks}. We also apply the method to find some equilibrium vortex-line
configurations in bulk superconductors and in films with a  square
lattice of columnar defects.  This system is of experimental interest
because there are several studies of vortices in films with a lattice
of columnar defects  created by lithographic techniques or by
bombardment with an electron beam \cite{harada,colf}. 

This paper is organized as follows. In Sec.\ \ref{sec.ldt} we discuss
the details of the theoretical model, numerical method and simulated annealing
algorithm. In Sec.\ \ref{sec.appl} we report the results of the above
mentioned applications. Finally in Sec.\ \ref{sec.concl}  we state our
conclusions.

\section{Theoretical model and numerical method}
\label{sec.ldt}

We consider a bulk uniaxially anisotropic superconductor with a
lattice of columnar defects (CD). 
We assume that  both the  vortex lines and the CD are straight and
parallel to the c-axis, and that the vortex lines
form a periodic VLL. Our problem  
is then to determine, for a given external field 
parallel to the c-direction, $H$, the VLL $primitive$ unit
cell, the number of vortex lines  and their positions within
this cell, that give the absolute minimum of the Gibbs free energy per
volume 
\begin{equation}
G=E_{\rm vv}+E_{\rm v-cd}-\frac{1}{4\pi}B(H-H_{c1})\; ,
\label{eq.gfe}
\end{equation}
where $E_{\rm vv}$ is the vortex line-vortex line {\it interaction} energy,
$E_{\rm v-cd}$ 
is the vortex-CD-lattice interaction energy, $B$ is the magnetic induction, 
 $H_{c1}=4\pi \epsilon/\phi_0$ is the lower critical 
field, and  \linebreak
$\epsilon=(\phi_0^2/16\pi^2\lambda^2_{ab})\ln{(\lambda_{ab}/\xi})$ 
is the vortex line self-energy
($\lambda_{ab} =$ penetration depth for currents parallel to the
ab-plane, $\xi=$ vortex core radius).

We first solve the simpler problem of finding the VLL with a
{\em fixed} number of vortex lines per unit cell, $n_v$, that minimizes $G$.
To find  which one of these  gives the absolute minimum  of $G$ we have 
to compare the free energies of the VLL obtained for the
same  $H$ and different $n_v$, if these are distinct.

Assuming that the VLL unit cell is  defined by the
vectors ${\bf L}_1$ and   ${\bf L}_2$, as shown in Fig.\ \ref{fig.ucel},
 or by the corresponding reciprocal lattice vectors ${\bf G}$, the 
London approximation  expression for $E_{\rm vv}$ is  
\begin{eqnarray}
E_{\rm vv}& = & \frac{\phi^2_0}{8\pi}\left[\sum^{n_v}_{\alpha,\, \beta=1}
\frac{1}{A_c}\sum_{\bf G}
\frac{e^{i{\bf G}\cdot({\bf x}_{\alpha}-{\bf x}_{\beta})}}
{1+\lambda^{2}_{ab}{\bf G}^2}\right.\nonumber\\
& &\left.-n_v \int\frac{d^2{\bf k}}{(2\pi)^2}\frac{1}{1+\lambda^{2}_{ab}{\bf
k}^2}\right] \; , 
\label{eq.evv}
\end{eqnarray}
where  $A_c$ is the unit cell area and ${\bf x}_{\alpha}$ ($\alpha
=1,...,n_v$) denotes the vortex 
lines positions within the unit cell. For the VLL shown in
Fig.\ \ref{fig.ucel},  $B=n_v \phi_0/A_c$ and  $A_c=L_1L_2\sin{\phi}$.

The vortex-line CD-lattice interaction energy  is assumed to be
the sum of the single vortex-single  CD interaction energies, namely
\begin{equation}
E_{\rm v-cd}=\frac{1}{NA_c}\sum^N_{j=1}\sum^{n_v}_{\alpha=1}\sum_{{\bf R}_d}
 U_{\rm v-cd}({\bf R}_j + {\bf x}_{\alpha}-{\bf R}_d) \; ,
\label{eq.evcd}
\end{equation}
where $N$ is the  number of vortex-lattice unit cells, ${\bf R}_j$
denotes the cell positions, ${\bf R}_d$ denotes the CD positions and
$U_{\rm v-cd}$ is the single vortex- single CD interaction energy per
unit length. The validity of  Eq.\ (\ref {eq.evcd}) requires that the
CD radius  $R$ be such that $R\sim 2\xi$, so that only a single vortex
line can be trapped by the CD, and that the vortex lines mean
separation is large  compared to $R$ \cite{cdlon}. We use for
$U_{\rm v-cd}$ an approximation based on the London theory results
derived in Ref:\cite{cdlon}, namely
\begin{eqnarray}
U_{\rm v-cd}(r) & = & \epsilon \, 
\ln{\left[1-\frac{R^2}{r^2}\right]} \; \;  \mbox{ $(r>2R)$}\\ 
U_{\rm v-cd}(r) & = & \epsilon
\left(2-\frac{r}{2R}\right)\ln{\frac{3}{4}} \; \;   \mbox{ $(r\leq 2R)$} \; . 
\end{eqnarray}
This expression interpolates smoothly between the large $r$ behavior
and the value of the energy per unit length of a vortex line pinned  by
a CD $\sim \epsilon\ln{(\sqrt{2}\xi/R})$, for $R\sim \sqrt{2}\xi$
\cite{cdlon}.

In order to study both bulk samples and films of thickness $D<\lambda_{ab}$
in the above described  framework, 
we use in the expression for $E_{\rm vv}$, Eq.\ (\ref {eq.evv}), instead
of $\lambda_{ab}$ an effective penetration depth $\Lambda\sim
\lambda^2_{ab}/D$. This mimics the dominant logarithmic 
vortex-vortex interactions at  distances  short compared to the film
screening length. The vortex-CD interaction is assumed to
be the same for bulk and films.

To numerically minimize  $G$, Eq.\ (\ref {eq.gfe}),   it is necessary to
evaluate $E_{\rm vv}$, Eq.\ (\ref {eq.evv}). Because the lattice sum in
this expression converges very slowly, a numerical method based on it
requires considerable amount of 
computer time. Doria \cite{mmd} showed that the lattice sum in
Eq.\ (\ref {eq.evv}) can be transformed into a series that converges
much faster. We use this series in our calculation to generate an
efficient numerical algorithm.   

Our minimization procedure, based on standard simulated annealing and
Monte Carlo techniques\cite{sma}, is as follows. For a given $H$ and
$n_v$, we  start with a chosen  unit cell shape and vortex lines
located within it. By moving the vortex lines, one at a time, and
deforming the unit cell continuously  we generate, using a Metropolis
algorithm, the equilibrium configurations of the cell shapes and vortex
lines positions within it, for a given fictitious temperature $T$. The
changes in $G$ caused by vortex line motion or by  unit cell
deformation are calculated using Doria's fast convergent series for
$E_{\rm vv}$,  and  a direct summation over a large CD-lattice for
$E_{\rm v-cd}$. The VLL that minimizes $G$ is identified with the
configuration at very low $T$. In order to reduce trapping in
metastable states, $T$ is cycled appropriately.

The VLL found by the above described method can be
either commensurate or incommensurate with the 
CD-lattice. Only the commensurate ones  correspond to
true minima of the Gibbs free-energy, since we restrict our search for
the $G$ minima to
periodic vortex-line arrangements. The incommensurate VLL, being
periodic, feel only the space average of the vortex-CD-lattice
potential, that is, a constant \cite{cminc}. Consequently, these  are
triangular VLL  that 
minimize $G$, Eq.\ (\ref {eq.gfe}), in the absence of the vortex-CD-
lattice interaction.      
As discussed in detail in  Sec.\ \ref{sec.racd},  we expect that in
situations where vortex-vortex interactions are 
strong, such as the ones considered here, our method is capable of
predicting  not only the  commensurate phases, but also  of giving
reasonable estimates for the  range of $H$ values over which they
minimize $G$.

\section{Applications}
\label{sec.appl}
In this section we apply the method described above to two physical
problems.

\subsection{Defect-free superconductor} 
\label{sec.dfs}

To illustrate the accuracy of our method, we first apply it to the well
known problem of vortex lines in a defect-free superconductor subjected
to an external magnetic field $H$ pointing in the c-direction. In this
case $G$ is minimized by  a triangular VLL with a single
vortex line in the primitive unit
cell. We minimize $G$ for several $H$ assuming $n_v=1$
and $n_v=2$.  We find the triangular lattice in all  cases considered.
For a given $H$, we compare the minima obtained for $n_v=1$ and
$n_v=2$.  We find for $n_v=2$ a unit cell that has twice the area as that  for
$n_v=1$  and  a value of $G$ that differs from that for $n_v=1$ by
less than 3 parts in $10^6$.  We also find that $B$, calculated by our
method, is very accurate when compared  with the known analytic
expressions for  $H\sim H_{c1}$ and for $H_{c1}\ll H \ll H_{c2}$
\cite{parks}. This is illustrated in Fig.\ \ref{fig.bhc} for the
$M\times H$ curve, instead of the $B\times H$ because $B\sim H$. The
errors seen in this figure indicate that our $B$ values differ from the
theoretical ones by less than 1\%.

\subsection{Regular array of columnar defects}
\label{sec.racd}

Now we consider  a superconductor  with a CD-lattice. 

We assume that the CD radius is $R=2\xi=0.02\lambda_{ab}$ and that
the CD-lattice 
is square with lattice constant  $a_{cd}=0.6\lambda_{ab}$. For films we
assume that $\Lambda=5\lambda_{ab}$. 

First we look for the matching phase for which
$B_{\phi}=\phi_0/a^{2}_{cd}$ and the VLL is square and
commensurate with the CD-lattice, with one vortex-line 
pinned  at each CD. We find that for both bulk and films with $n_v=1$
the matching phase minimizes $G$ in the  range of $H$ values
$H^{<}_{\phi}<H<H^{>}_{\phi}$. We obtain for bulk
$H^{<}_{\phi}= 1.01B_{\phi}$, $H^{>}_{\phi}=1.10B_{\phi}$, and for
films  $H^{<}_{\phi}=H^{>}_{\phi}= 1.00B_{\phi}$. When $H$ is just
outside this range  we find that a VLL incommensurate with
the CD-lattice minimizes $G$. The $H$-range for films is smaller than
that for bulk due to the  value of $\Lambda=5\lambda_{ab}$.

We recall that by our method  $H^{<}_{\phi}$ and
$H^{>}_{\phi}$ are the boundaries in the $B\times H$ phase diagram
where the matching state becomes unstable with respect to an
incommensurate triangular VLL. We may ask what is the
relationship between these boundaries and true ones obtained if $G$ were
minimized without the restrictions imposed by our method. To answer
this question it is necessary to guess the  state to which the matching
phase would become unstable to at these boundaries.  One possibility is
that the matching state becomes unstable with respect to the state
obtained from it by the addition  of
vacancies or interstitials \cite{blam}. Another possibility is that at
these boundaries a commensurate-incommensurate (CI) transition takes place
\cite{cminc}. We believe that the former is possible only if the vortex-CD
interaction is stronger than vortex-vortex interactions, which is not
the case here. In our calculation the parameters
  are such that these two interactions are of
comparable strength so, we believe, that the true boundaries here
correspond to a  CI transition. It is known that for strong
vortex-vortex interactions, and not too close to the CI transition
point, the incommensurate state differs from 
the triangular lattice  by small incommensurate elastic distortions
\cite{cminc}. Thus, we believe that our method makes a reasonable
approximation for the incommensurate state (except close to the CI
transition), and that the  estimates obtained from it for the matching 
state phase boundaries are close to the true ones.

We also apply our method with $n_v=2$ a value of $H$ in
the range $H^{<}_{\phi}<H<H^{>}_{\phi}$. We find that the matching
state with a rectangular unit cell twice as large as that for $n_v=1$
minimizes $G$ and that the  $G$-value differs from that for $n_v=1$ by
less than 1 part in $10^4$.    
 
Next we study VLL commensurate with the CD-lattice for
$B< B_{\phi}$. These  can be characterized  as follows. We assume that
the CD-lattice primitive unit cell 
is a square  of side $a_{cd}$ with unit vectors ${\bf e}_x$ and 
${\bf e}_y$ oriented along the sides. We expect that for these VLL all 
vortex lines are pinned by the CD. Thus, the VLL  primitive
unit cell vectors, ${\bf a}_1$ and ${\bf a}_2$ can be written as
\begin{equation}
{\bf a}_i= a_{cd}(n_{ix}{\bf e}_x + n_{iy}{\bf e}_y)\; (i=i,2)\;.
\label{eq.pucv}
\end{equation}
If there are  $n_v$  vortices in the unit cell, the magnetic induction
in this state is  
\begin{equation}
B_c=B_{\phi}\frac{n_v}{\mid n_{1x}n_{2y}-n_{1y}n_{2x}\mid} \; '
\label{eq.bc}
\end{equation}
where $n_v$ is restricted by $n_v<\mid n_{1x}n_{2y}-n_{1y}n_{2x}\mid$.
Thus, when $B/B_{\phi}$ is rational,  there are many non-equivalent
VLL satisfying the above stated conditions. We apply our numerical method 
to  determine which one of these minimize $G$, and to estimate the range
of  $H$ values over which this minimum is stable. 

We study   $B=B_{\phi}/5$, $
B_{\phi}/4$, $B_{\phi}/3$, $B_{\phi}/2$, $2B_{\phi}/3$ 
with $n_v=1$ and  $B=2B_{\phi}/5 $ with $n_v=2$.
We find that for films  the commensurate
VLL  shown in Fig.\ \ref{fig.vll} minimize $G$ with $H=B$. The $H$
range where these minima are stable is around $H=B$ for which these VLL 
minimize $G$ is less than $10^{-3} B_{\phi}$.
For bulk  we find that for  $B_{\phi}/4$ and
$B_{\phi}/2$ the same VLL minimize $G$ for 
$H=0.330B_{\phi}$ and  $H=0.528B_{\phi}$, respectively. The $H$-range was not
investigated in detail in this case.  

  Our results for
$B=B_{\phi}/4$ and $B=B_{\phi}/2$ agree with recent Lorentz microscopy
experiments reported in Ref.\cite{harada}.

\section{Conclusions}
\label{sec.concl}

In conclusion then we develop a numerical method to calculate 
the zero-temperature phase diagram of a bulk superconductor in the
presence of an external field $H$ and show that it gives accurate
results for the $B\times H$ curve of an ideal superconductor and that
it can make predictions about the zero-temperature phase diagram of the
non-trivial problem of vortex lines in the presence of a CD-lattice.
 The method can be
extended to study the zero-temperature phase diagram for anisotropic
superconductors in tilted fields and to long cylinders of rectangular cross
sections in axial fields. Work along these lines is under way and will
be reported elsewhere.

\acknowledgments
We thank Prof. M. Doria for stimulating  conversations and helpful
suggestions.
This work was supported in part by CNPq-Bras\'{\i}lia/Brazil and FAPERJ.

\epsfxsize= .9\hsize  \vskip 1.5\baselineskip
\centerline{ \epsffile{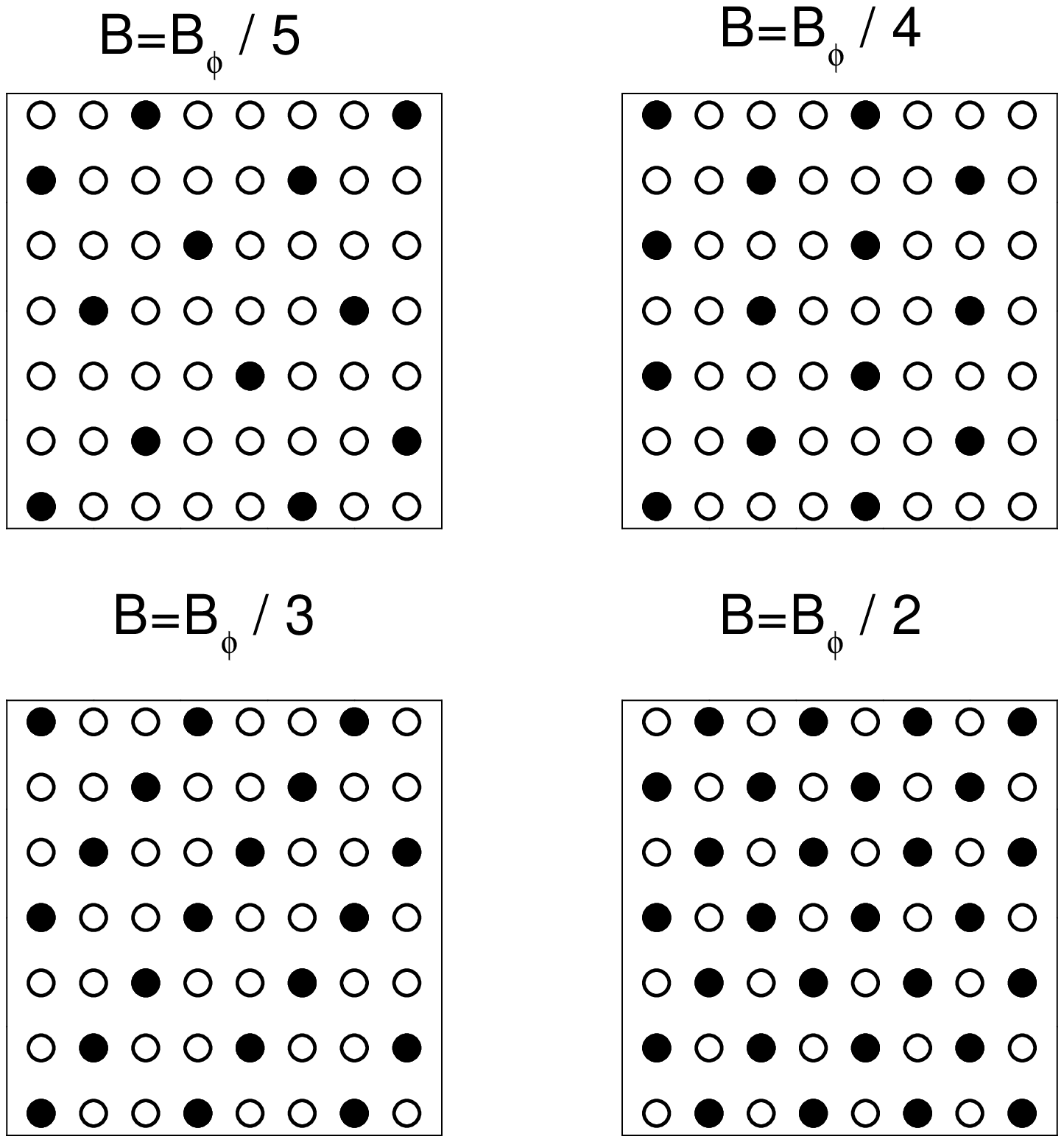}}
\begin{figure}
\caption{Vortex-line-lattice unit cell with $n_v=3$.}
\label{fig.ucel}
\end{figure}

\newpage
\epsfxsize= .9\hsize  \vskip 1.5\baselineskip
\centerline{ \epsffile{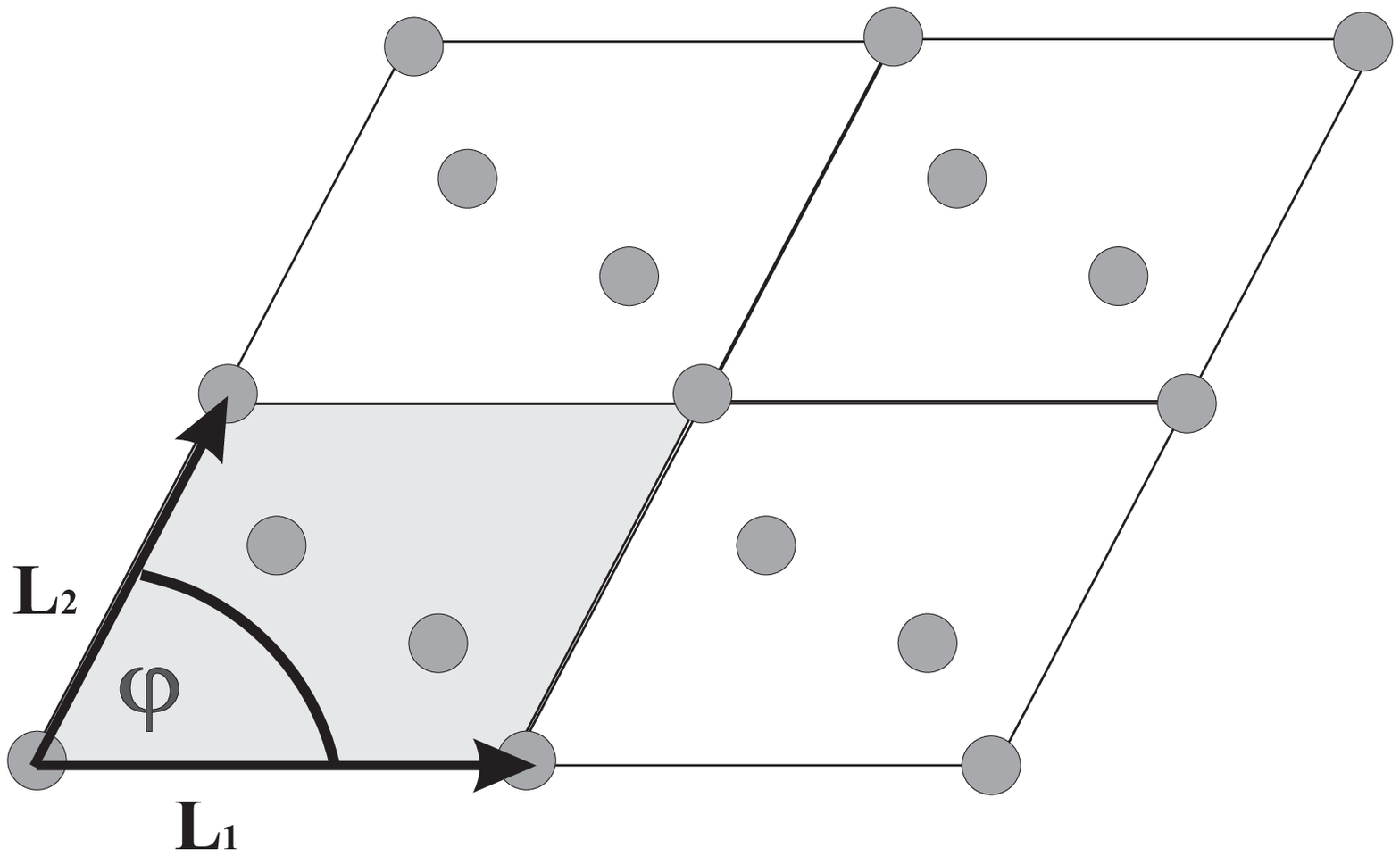}}
\begin{figure}
\caption{Magnetization for a defect free superconductor versus $H$
compared to theoretical results for low $H$ ( full curve) and
intermediate $H$ ( dashed curve). }
\label{fig.bhc}
\end{figure}

\epsfxsize= .9\hsize  \vskip 1.5\baselineskip
\centerline{ \epsffile{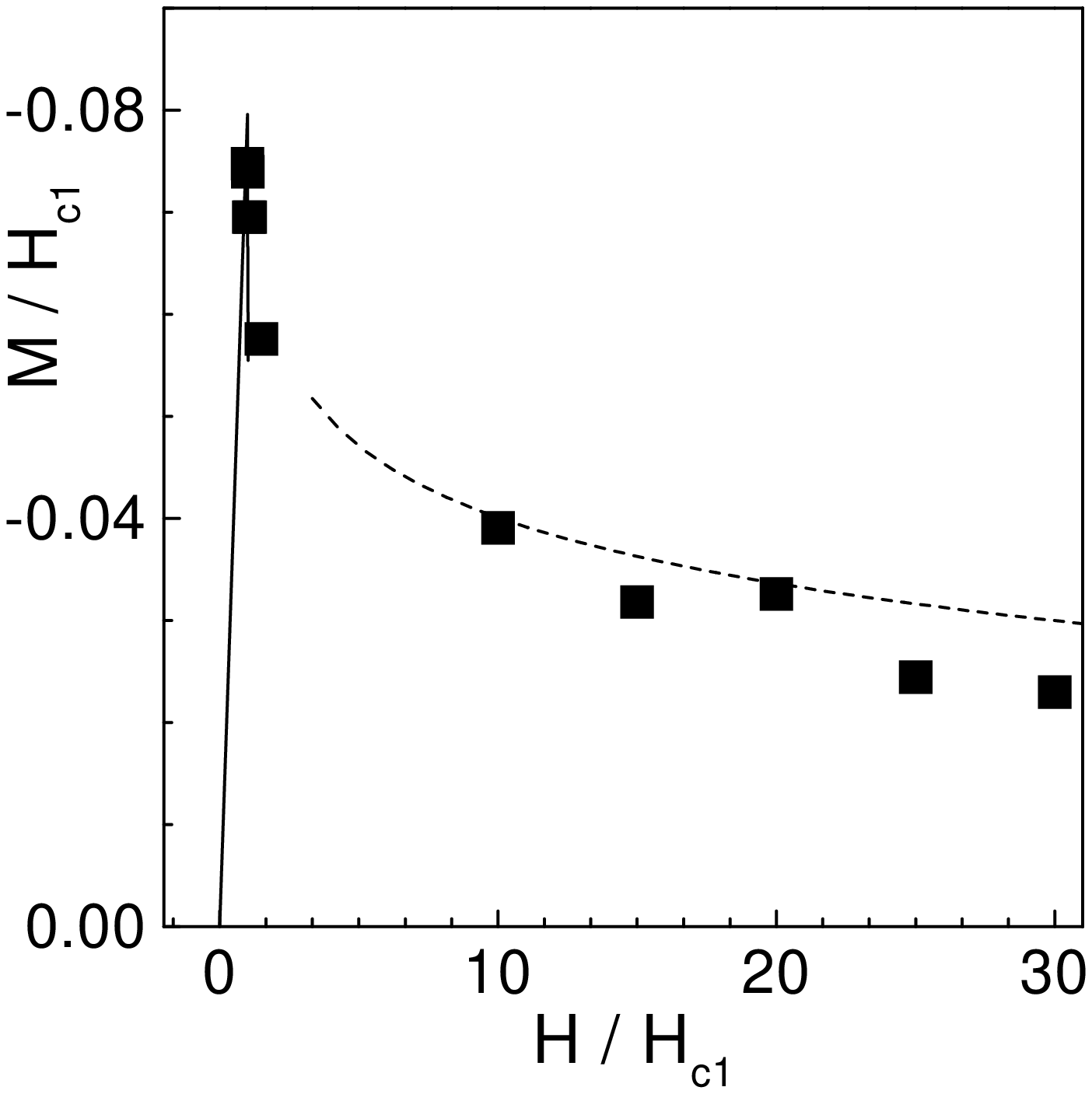}}
\begin{figure}
\caption{Commensurate vortex-line-lattices that minimize the Gibbs 
free-energy.} 
\label{fig.vll}
\end{figure}

\end{multicols}
\end{document}